\documentclass[pss]{wiley2sp} 
\usepackage{amsmath}
\usepackage{bm}        
\usepackage{w-greek}
\usepackage{graphicx}      
\usepackage{epstopdf}
\usepackage{dcolumn}  

\begin{document}

\title{Impurity states on the honeycomb lattice using the Green's function method}

\titlerunning{Impurity states on honeycomb lattice }

\author{%
  Mohammad Sherafati\textsuperscript{\Ast},
  Sashi Satpathy }

\authorrunning{M. Sherafati et al.}

\mail{e-mail
  \textsf{ms9n9@mail.mizzou.edu}, Phone:
  +1 573 884 2204, Fax: +1 573 882 4195}

\institute{%
Department of Physics $\&$ Astronomy, University of Missouri,
Columbia, MO 65211, USA \\}

\received{XXXX, revised XXXX, accepted XXXX} 
\published{XXXX} 


\abstract{%
\abstcol{%
Using the Green's function method, we study the effect of an impurity  potential on the electronic structure of the honeycomb lattice in the one-band tight-binding model that contains both
the nearest neighbor ($t$) and the  second neighbor ($t'$) interactions.
The model is relevant to the case of the substitutional vacancy in graphene.
If the second neighbor interaction is large enough ($t' > t /3$), then the linear Dirac bands no longer occur at the Fermi energy and the electronic structure is therefore fundamentally changed.
With only the nearest neighbor interactions present, there is particle-hole symmetry, as a result of which  the
vacancy induces a  ``zero-mode" state
}{
at the band center with its
wave function entirely on the majority sublattice, i. e., on the sublattice not containing the vacancy.
With the introduction of the second neighbor interaction, the zero-mode state broadens into a resonance peak and its wave function spreads into both sublattices, as may be argued from the Lippmann-Schwinger equation. The zero-mode state disappears entirely for the triangular lattice and if $t'$ is large for the honeycomb lattice as well. In case of graphene, $t'$ is relatively small, so that a well-defined zero-mode state  occurs in the vicinity of the band center.
}}

\maketitle

\section{Introduction}

There is considerable current interest on the behavior of graphene, which is a two-dimensional honeycomb lattice of carbon atoms.   The honeycomb lattice results in the unusual linearly-dispersive band structure,  which leads to a host of interesting properties directly originating from the linearity of the band structure such as Klein tunnelling, Zitterbewegung, novel Quantum Hall effect, and tabletop quantum electrodynamics   \cite{Review1,Review2}.
In this paper, we study the impurity states, which also shows unusual behaviors originating from the linearity of the band structure and the particle-hole symmetry present in the bipartite honeycomb lattice.
Since the physical properties of a material are often controlled by the presence of impurities, the study of the single vacancy forms the basic foundation  for the understanding of the behavior of more complex defects.

There have been several theoretical studies of the vacancy center in the honeycomb lattice using the tight-binding model   \cite{castroneto,hjort,Ranjit-defect}. But, they are either restricted to the nearest neighbor ($NN$) interactions only or they use finite-size supercell calculations. In the latter case, it becomes difficult to separate the effects of the finite size from the effects of the presence of the impurity. One interesting property, however, has been clearly established in the form of a theorem, viz., the presence of the zero-mode state, which states that if only $NN$ interactions are present in a bipartite lattice, then the substitutional vacancy produces a ``zero-mode" state with energy equal to the on-site energy, with its wave function living entirely on the sublattice opposite  to that of  the vacancy  \cite{castroneto,Ranjit-defect}.  In addition, the zero-mode state is quasilocalized with the wave function falling off as $1/r$ with distance, in the approximation of the linear band structure. However, the effect of the absence of the  particle-hole symmetry, which is the case if higher-neighbor interactions are present, on the zero-mode state has not been adequately studied. In this paper, we study in detail the effect of the  second neighbor ($2NN$) interaction  on the zero-mode state. If the $2NN$ interaction is strong, then the honeycomb lattice shows the behavior of the triangular lattice, where the zero-mode state disappears. The presence of the $2NN$ interaction breaks the particle-hole symmetry in a bipartite lattice and the inclusion of the further neighbor interactions do not change the physics in any essential ways.

The method we will use in this paper is the Green's function (GF) approach, which has been extensively used for  defects in solids and has been pioneered by Sankey and coworkers  \cite{Sankey} for the study of defects in semiconductors.

\section{Tight-binding electronic structure}
\begin{figure}
\centering
\begin{minipage}{1.0cm}
\includegraphics[width=3.5cm]{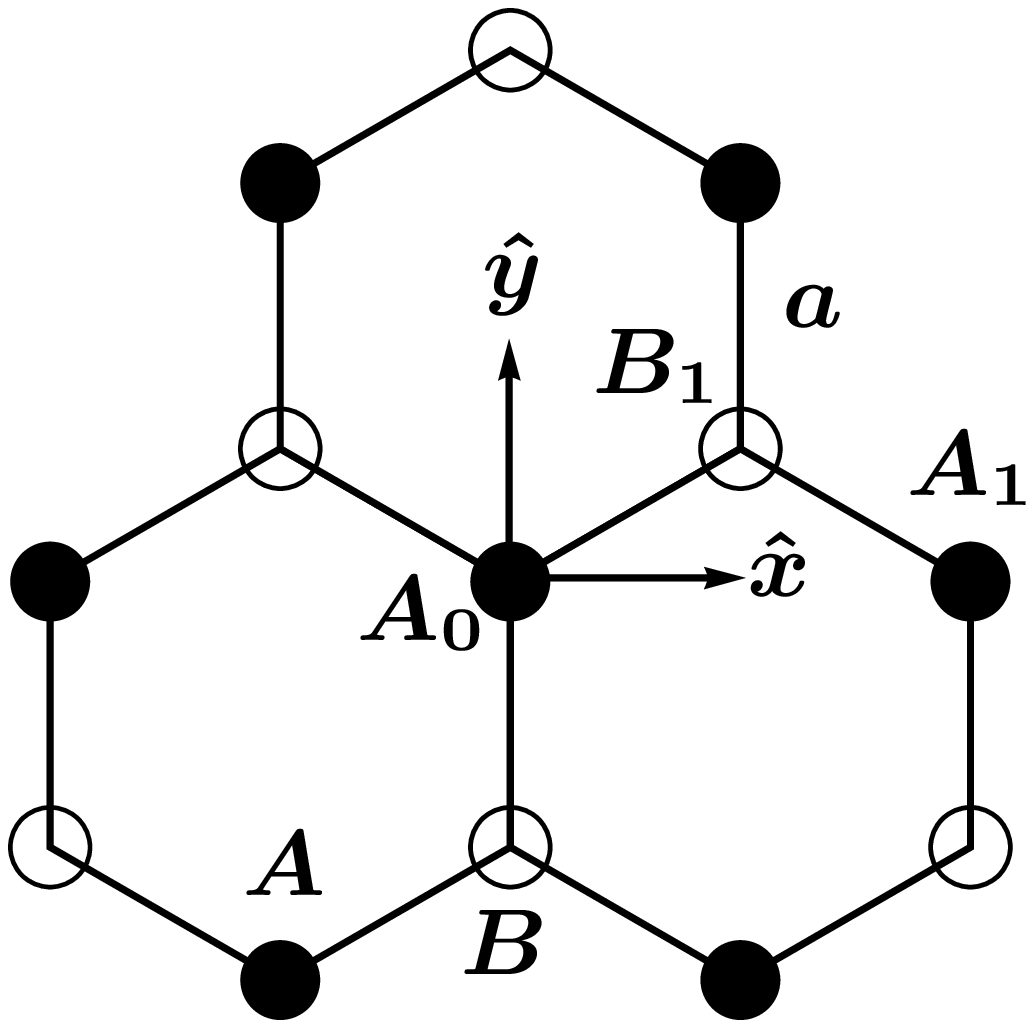}
\end{minipage}
\begin{minipage}{5.5cm}
\hspace{3cm}
\includegraphics[width=2.5cm]{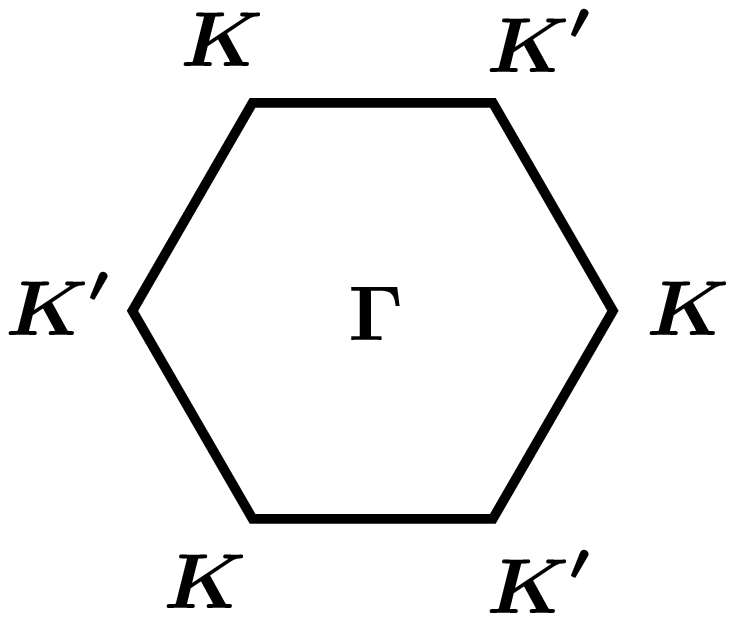}
\end{minipage}
\caption{The honeycomb lattice with two different sublattices, $A$ and $B$, shown as full and open circles and the corresponding Brillouin zone with the Dirac points $K$ and $K'$. Impurity is placed at the site  $A_0$ and two of its neighboring sites
are labelled as $A_1$ and $B_1$.
    }
\label{graphene}
\end{figure}

Honeycomb lattice is a two-dimensional  structure formed by two interpenetrating triangular lattices,  $A$ and $B$, with two atoms in the unit cell indicated in Fig. (\ref{graphene}). The tight-binding Hamiltonian is
\begin{equation}
{\cal H}_0=-t \sum_{NN} c^{\dagger}_{i\alpha}c_{j\beta} +t' \sum_{2NN} c^{\dagger}_{i\alpha}c_{j\beta},
\label{hamil}
\end{equation}
keeping only the $NN$  and the $2NN$ terms, where $c^{\dagger}_{i\alpha}, c_{i\alpha}$ are the creation and annihilation operators for the electrons with $i \alpha$ being the cell and sublattice index.
From density-functional calculations, the hopping parameters for graphene are $ t =  2.56$ eV and $t' = 0.160$ eV   \cite{Nanda-Graphene}.

The band structure is simple to obtain. Using the Bloch function basis   $|\bm{k}\alpha\rangle= N^{-1/2} \sum_i  e^ {i \bm{k}. \bm{r}_{i \alpha}}   | i \alpha\rangle$, where $\bm{r}_{i \alpha}={R}_i+\bm{\tau}_\alpha$ are the atom positions and $N$ is the number of unit cells in the crystal, the Hamiltonian becomes

\begin{figure}[h]%
\centering
\includegraphics[width=7.0cm]{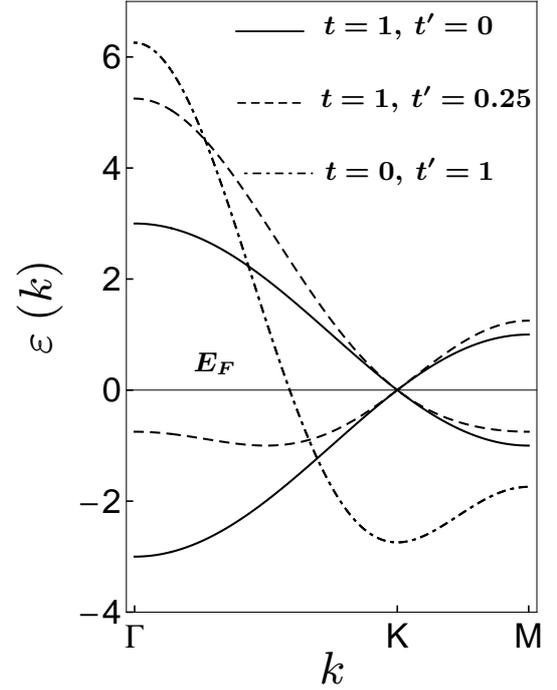}
\caption{%
The band structure of the honeycomb lattice for different values of the hopping parameters. For $t' > t/3$, the linearly-dispersive Dirac band structure occurring at $E_F$ for the half-filled system is lost. With $t'/t \rightarrow \infty$ the honeycomb lattice becomes effectively two decoupled triangular sublattices, whose band structure is shown by the dash-dotted line. The Fermi energy is indicated in the figure for the half-filled system (one electron per site).
}
\label{band}
\end{figure}
\begin{equation}
{\cal H}_0=
 \left( \begin{array}{cc}
f'(\bm{k}) &  f(\bm{k})\\
f^*(\bm{k}) & f'(\bm{k})
\end{array}\right),
\label{H0}
\end{equation}
with $f(\bm{k}) = -t \ \sum_{j=1}^{3} e^{i \bm{k} \cdot  \bm{d_j}} = -t [2 \exp (ik_ya/2) \cos  \\ ( \sqrt 3 k_xa/2)+ \exp(-ik_ya)]$ and
 $f'(\bm{k}) = t' \ \sum_{j=1}^{6} e^{i  \bm{k} \cdot  \bm{d_j}} = t' [2 \cos (\sqrt3k_xa)+4 \cos(\sqrt 3 k_xa/2)
 \cos( 3 k_ya/2)]$,
 where $\bm{d_i}$'s are  the $NN$ positions for $f(\bm{k})$ and the $2NN$ positions for  $f'(\bm{k})$. These two functions happen to be related:
$f'(\bm{k})   =     t' |f(\bm{k})|   ^2/t^2     -3 t'$, so that the eigenstates are given by
\begin{eqnarray}
\varepsilon_\pm  (\bm{k})
                               &=&  \pm|f(\bm{k})|+t'  t^{-2}   |f(\bm{k})|^2     -3t'  \nonumber  \\
\Psi_{\bm{k}\pm} ^0 &=& \frac{1}{\sqrt 2}
 \left(
 \begin{array}{c}
\pm e^{i \theta_{\bm {k}}    } \\ 1
\end{array}
\right),
\label{estate}
\end{eqnarray}
where $\pm$ denotes the conduction and the valence bands and the phase factor $e^{i \theta_{\bm{k}}  }= f(\bm{k}) /   |f(\bm{k})|$ justifies the interpretation of the wave function in terms of pseudo-spins.

The band structure is shown in Fig. (\ref{band}). The linear-dispersion near the Dirac points $K$ or $K'$, viz.,
$\varepsilon (q) = \pm v_F q - 3 t'$, apart from the unimportant energy shift, is clearly seen in the band structure and the Fermi velocity $v_F = 3 t a / 2$ does not depend on $t'$. However, if the $2NN$ interaction is strong, the bands cross the zero of energy at points other than the Dirac points thus moving the Fermi energy $E_F$ away from the linearly-dispersive Dirac points, so that the interesting physics originating from the Dirac points is lost. This happens when $t' \ge t/3$, so that in that case the band structure of the honeycomb lattice becomes fundamentally different, since there is no linearity of the band structure at the Fermi energy. The parameters for graphene are far  from this condition, so that the Dirac point physics continues to remain valid even under external perturbations such as strain or electric fields. For very large values of the $2NN$ interaction $t < < t'$, one recovers the band structure of the triangular lattice as the two sublattices become effectively decoupled.

\section{Green's functions and the impurity states}

The impurity is modeled by  adding an on-site perturbation $V$ to  the Hamiltonian
\begin{equation}
{\cal H}  =
{\cal H}_0+ V,
\label{hamil}
\end{equation}
where  $ V=U_0 c^{\dagger}_{0A}c_{0A} $, with the impurity located in the central cell on the $A$ sublattice  and $U_0$ being the strength of the impurity potential (vacancy corresponds to $U_0 \rightarrow \infty$).

The key quantity to compute is the full GF $G$, which is related to the unperturbed GF $G^0$ through the Dyson's equation $G = G^0 + G^0VG$. With the localized impurity potential, the Dyson's equation becomes
\begin{equation}
G_{i\alpha, j \beta}=G^0_{i\alpha, j \beta}+U_0 G^0_{i\alpha, 0 A}G_{0A, j \beta},
\label{Dyson-1}
\end{equation}
where $G^0_{i\alpha, j \beta} ( E) \equiv \langle i \alpha| G^0(E)|j \beta \rangle$. This equation may be inverted as usual to yield the full GF
\begin{equation}
G_{i\alpha, i\alpha}=G^0_{i\alpha, i\alpha} + \frac{U_0 G^0_{i\alpha,0A}  G^0_{0A, i\alpha} }  {1-U_0G^0_{0A, 0 A}}.
\label{GAA}
\end{equation}

To calculate the real-space unperturbed GF,  we first express it in terms of the momentum GF  $G^0_{\alpha \beta} (\bm{k},E) \equiv
\langle \bm{k} \alpha| G^0(E)|\bm{k} \beta \rangle $. Using the expression
\begin{equation}
G^0(E) = (E+i \eta - {\cal H}_0)^{-1}
= \sum_{\bm{k}s} \frac{|\Psi^0_{\bm{k}s}\rangle \langle\Psi^0_{\bm{k}s}|}{E+i \eta - {\cal H}_0},
\label{G0}
\end{equation}
where $s = \pm$ is the band index, and taking the real-space matrix elements, we find
\begin{eqnarray}
G^0_{i\alpha, j \beta} ( E) &=& \langle i \alpha| G^0(E)|j \beta \rangle  \nonumber \\
&=&\frac{1}{\Omega_{BZ}}\int d^2\bm{k} \ e^{i \bm{k} \cdot (\bm{r}_{i \alpha}-\bm{r}_{j \beta})}G^0_{\alpha \beta} (\bm{k},E),
\label{GF-Site}
\end{eqnarray}
where
\begin{align}
G^0_{\alpha \beta} ( \bm{k},E) = \mu_{\bm{k},E} \left(
\begin{array}{cc}
E+i\eta-f'(\bm{k}) &  f(\bm{k})\\
f^*(\bm{k}) & E+i\eta-f'(\bm{k})
\end{array}
\right),
\label{GF-Mom}
\end{align}
with $\mu_{\bm{k},E}=[(E+i\eta-f'(\bm{k}))^2-|f(\bm{k})|^2]^{-1} $. Thus the calculation of $G^0_{i\alpha, j \beta}$ boils down to the numerical integration in Eq. (\ref{GF-Site}) and the full GF with the impurity is obtained from Eq. (\ref{GAA}). The faster Horiguchi method may be used for the evaluation of the unperturbed GF for the case with $NN$ hopping only \cite{Horiguchi,M-RKKY}.

\begin{figure}[htp]%
\includegraphics[scale=0.6]{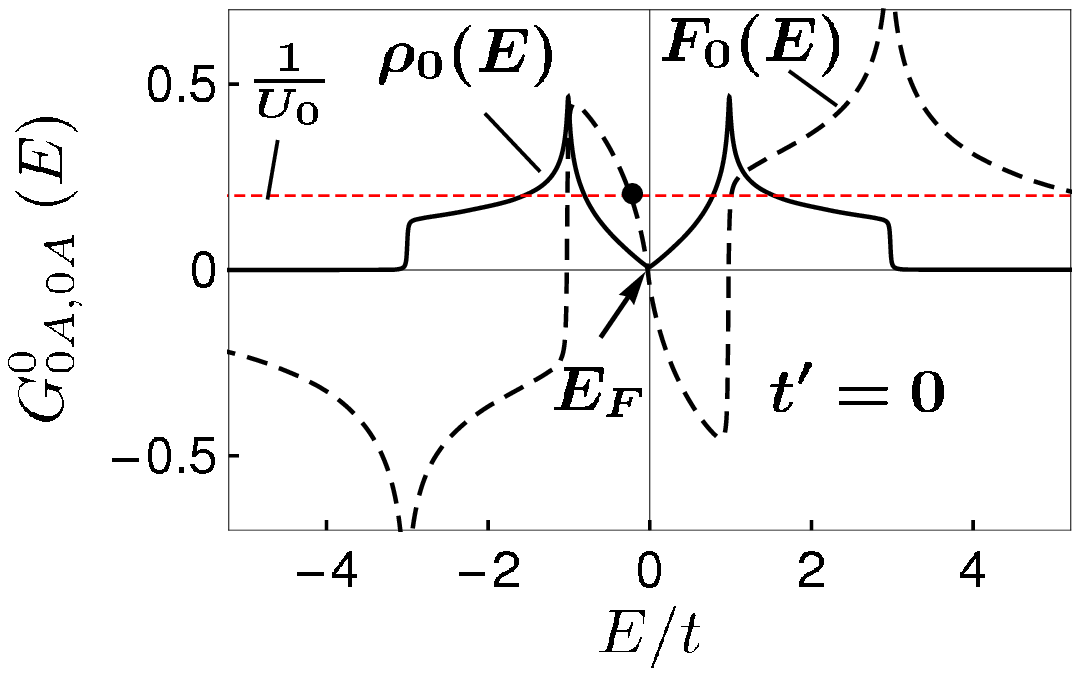}
\includegraphics[scale=0.6]{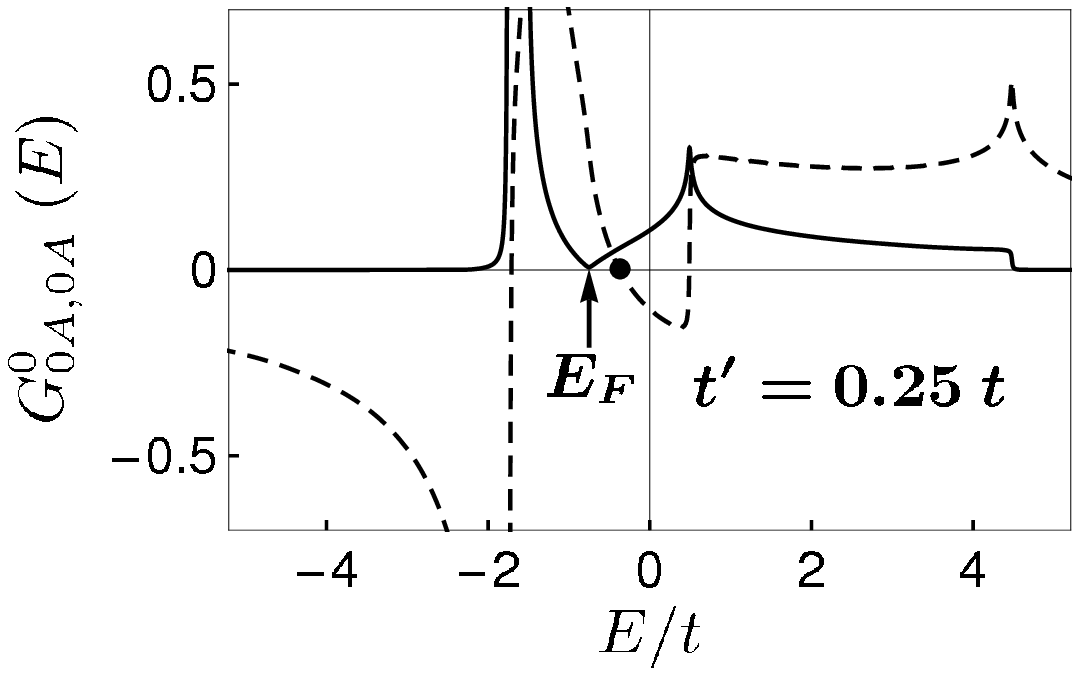}
\includegraphics[scale=0.6]{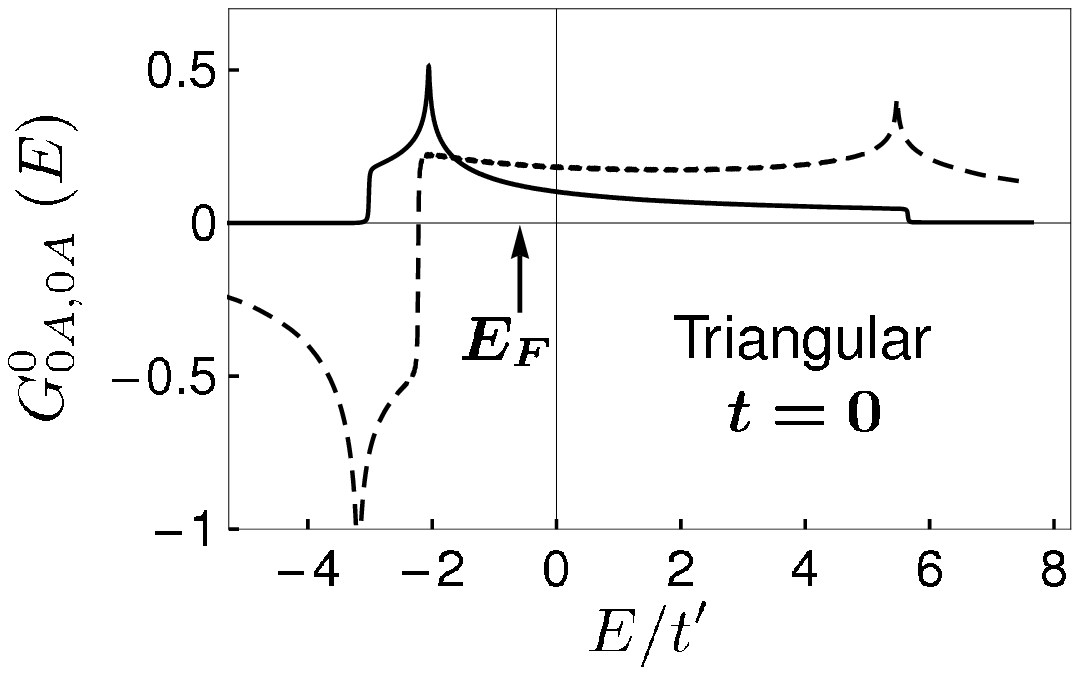}
\caption{%
Impurity state from the Dyson's equation. Dashed and solid lines indicate respectively the real  and  imaginary parts -- the latter multiplied by $-\pi^{-1}$ to yield the DOS $\rho_0 (E)$ --  of the unperturbed GF for three different cases:
(a) $ t =1, t' =0$ (top), (b) $ t =1, t' =0.25$ (middle), and (c) $ t =0, t' = 1$ (bottom). The last case corresponds to two decoupled triangular lattices.
In the top two figures, black dots show the energy of the impurity state with $U_0/ |t| = 5$ (top) and
$U_0 = \infty$ (middle). For $U_0 = \infty$, corresponding to the vacancy, the impurity state occurs exactly at $E_F$ in the top figure due to particle-hole symmetry, while in the middle figure it is displaced from $E_F$ due to lack of symmetry, and in the bottom figure the corresponding impurity state has disappeared.
}
\label{Fig-Dyson}
\end{figure}

 All quantities of interest may be expressed in terms  of the full GF, e. g., the local density of states (LDOS) at a specific site is expressed as
$\rho_{i\alpha} (E) = - \pi^{-1} {\rm Im} \ G_{i\alpha, i\alpha}(E)$.
From Eq. (\ref{GAA}), the LDOS at the impurity site has an especially simple form
\begin{equation}
\rho_{0A }(E)= \frac{\rho_0(E)}{(1-U_0 F_0(E))^2+(\pi U_0\rho_0(E))^2},
\label{LDOSAA00}
\end{equation}
where $\rho_0 (E) = -\pi^{-1} {\rm Im} \ G^0_{0A, 0A}(E) $ is the unperturbed DOS, which is the same for every site, and $F_0(E)\equiv {\rm Re} \  G^0_{0A, 0A}(E)$.
According to Eqs. (\ref{GAA}) and (\ref{LDOSAA00}), the contribution of the impurity has a sharp peak at the resonance energy $E_0$ satisfying the resonance condition
\begin{equation}
1-U_0F_0(E_0)=0.
\label{E0}
\end{equation}

The sublattice-projected densities of states $\rho_A(E)$ or $\rho_B(E)$ may be obtained by taking the trace of the GF:
\begin{equation}
\rho_\alpha (E) = \sum_i G_{i\alpha, i\alpha} (E)
\end{equation}
and this can be expressed in terms of the central-site GF in real and momentum space, $G^0_{0A, 0A}(E)$ and
$G^0_{AA} (\bm{k}, E)$, respectively, as shown in our earlier work   \cite{Ranjit-defect}.
The expression for the total DOS is obtained from the sum $\rho (E) = \rho_A (E) + \rho_B (E)$ to yield
\begin{equation}
\rho(E)=2\rho_0(E)+\frac{1}{N} \times
\frac{- U_0 [U_0 \rho_0(E)F'_0(E)+\rho'_0(E)(1-U_0 F_0(E))]}
 {(1-U_0 F_0(E))^2+(\pi U_0\rho_0(E))^2},
\label{DOSTot}
\end{equation}
where the prime denotes the derivative.
%
\begin{figure}
\centering
\includegraphics[scale=0.6]{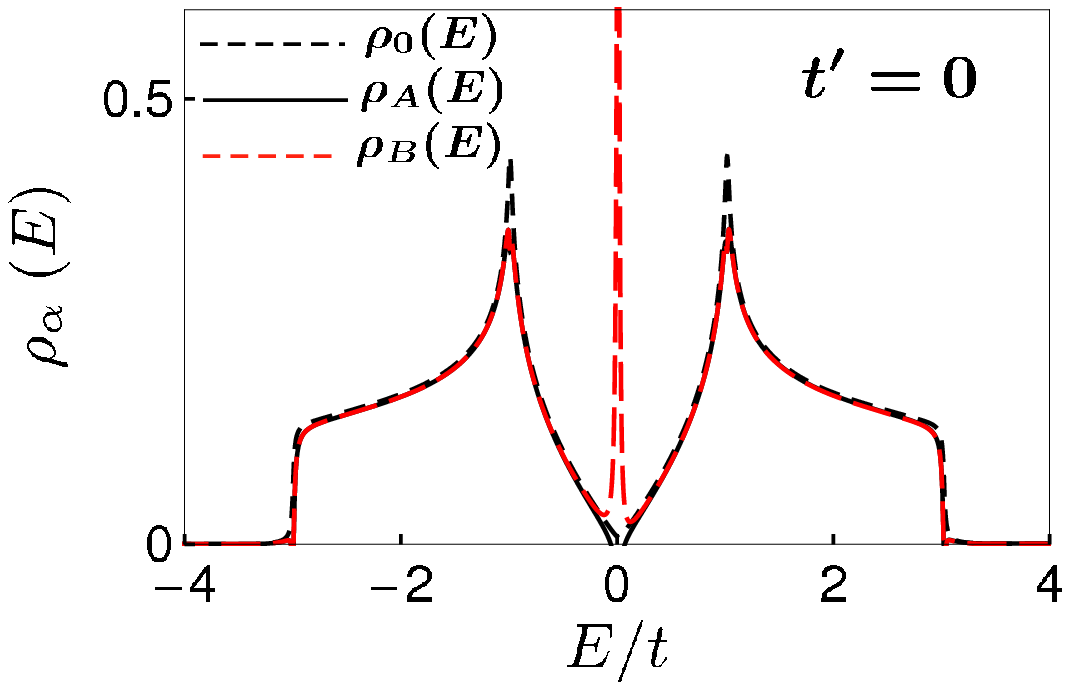}
\includegraphics[scale=0.6]{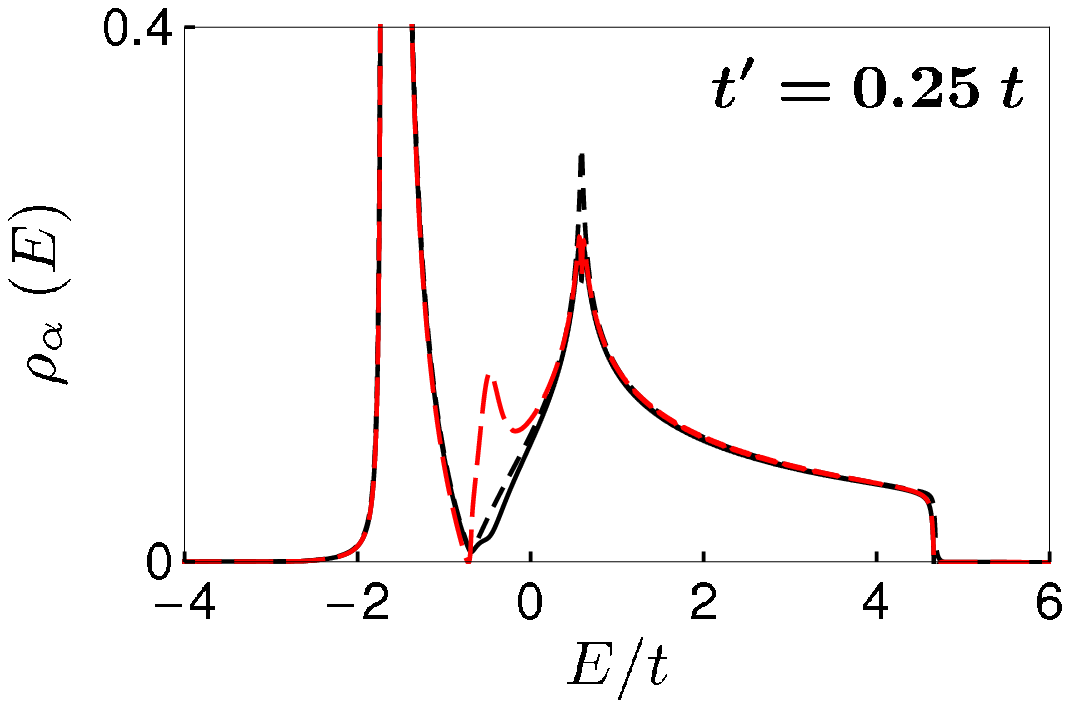}
\includegraphics[scale=0.6]{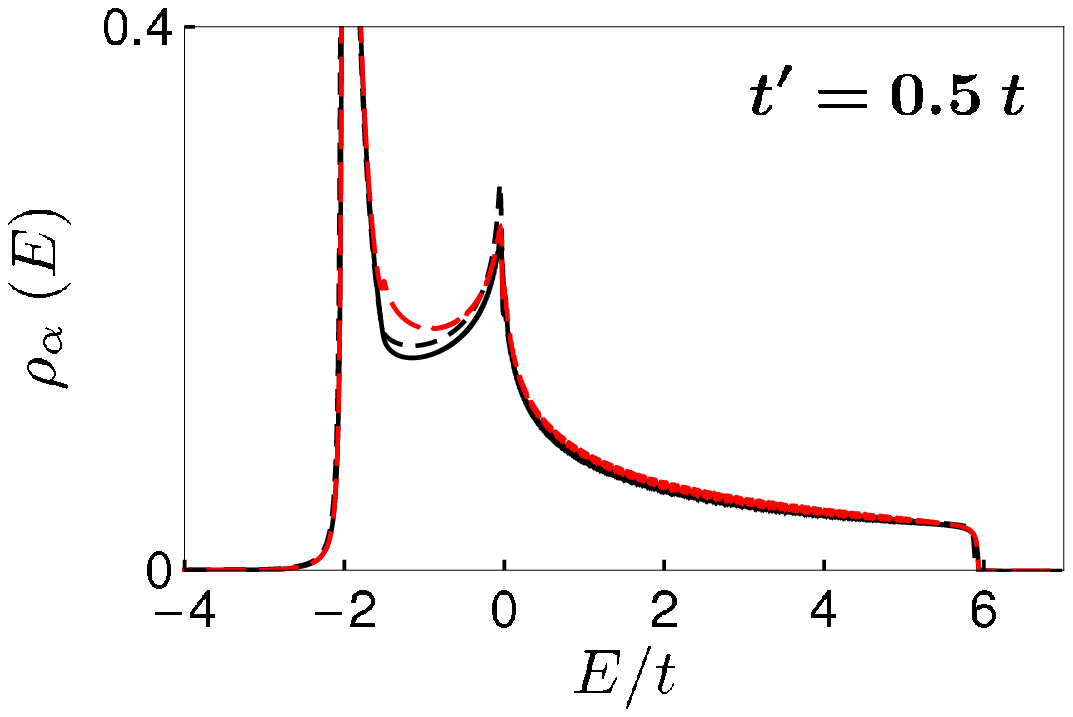}
\caption{ Sublattice densities of states (black solid and red dashed lines) together with the unperturbed DOS (black dashed line). With the presence of the $2NN$ interaction, the localized zero-mode state (top) turns into a
finite-width resonance state (middle), which disappears if $t'$ is large (bottom). The impurity potential was taken to be $U_0 / t = 100$ in the figure.
}
\label{sublattice-DOS}
\end{figure}

The graphical solution of Eq. (\ref{E0}) for several cases  are shown in Fig. (\ref{Fig-Dyson}).
Solution $E_0$ determines the resonance (inside the continuum band) and bound states (outside the continuum band) in the presence of the impurity. According to Eq. (\ref{LDOSAA00}), the solutions at the van Hove singularities will produce a negligible change in the LDOS, since either the  real or the imaginary part  of the unperturbed GF diverges at these points.

 Several points may be made from Fig. (\ref{Fig-Dyson}), where we have shown the evolution of the GF as we go from the honeycomb to the triangular lattice by changing the strength of $t'$:

  (i) The top figure, which shows the solution of the Dyson's equation with only the NN interaction, shows that  as the impurity potential $U_0$ is increased to infinity, the solution $E_0$ indicated by the black dot approaches the zero of energy resulting in the so-called ``zero-mode" state. The particle-hole symmetry demands that the real part of the GF $F_0 (E)$ is antisymmetric about the zero of energy, so that the resonance state occurs at $E_0 = 0$ just from symmetry. Because the unperturbed DOS $\rho_0 (E)$ is zero at the resonance, it results in a sharp resonance state of zero width as seen from Fig. \ref{sublattice-DOS} (top panel).

  (ii) With the second $2NN$ interaction present, particle-hole symmetry is no longer present, resulting in a DOS that is no longer symmetric about $E_F$, so that  the real part of the GF is not required to be antisymmetric. This can be seen from the relation ${\rm Re} \ G (r, E) = \pi^{-1} P \int_{-\infty}^\infty dE' (E' - E)^{-1}  {\rm Im} \ G (r, E')$. This is seen from the middle figure in Fig. \ref{Fig-Dyson}, where the solution $E_0$ occurs no longer at the band center, but is shifted from it, the sign of the shift depending on the relative signs of $t$ and $t'$.  The resonance peak now acquires a finite width being in the band continuum as seen from the middle part of Fig. \ref{sublattice-DOS}.

  (iii) As discussed in the previous section, if $t' > t/3$, the band structure is drastically changed with the Fermi energy no longer occurring at the Dirac points, even though the bands may be linear there. For somewhat larger value of $t'$, the zero-mode solution at the band center disappears altogether. This may be seen from the bottom parts of Figs. (\ref{Fig-Dyson}) and (\ref{sublattice-DOS}), which show the GF and the DOS, respectively, where $t'$ is substantially large.

\begin{figure}
\centering
\includegraphics[scale=0.6]{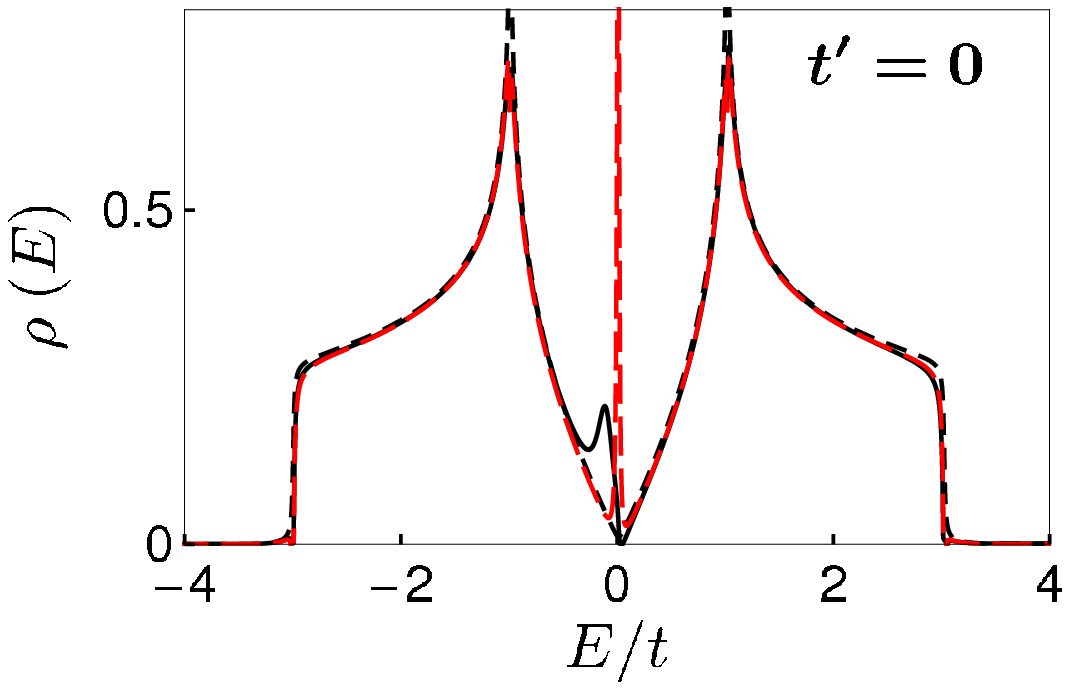}
\includegraphics[scale=0.6]{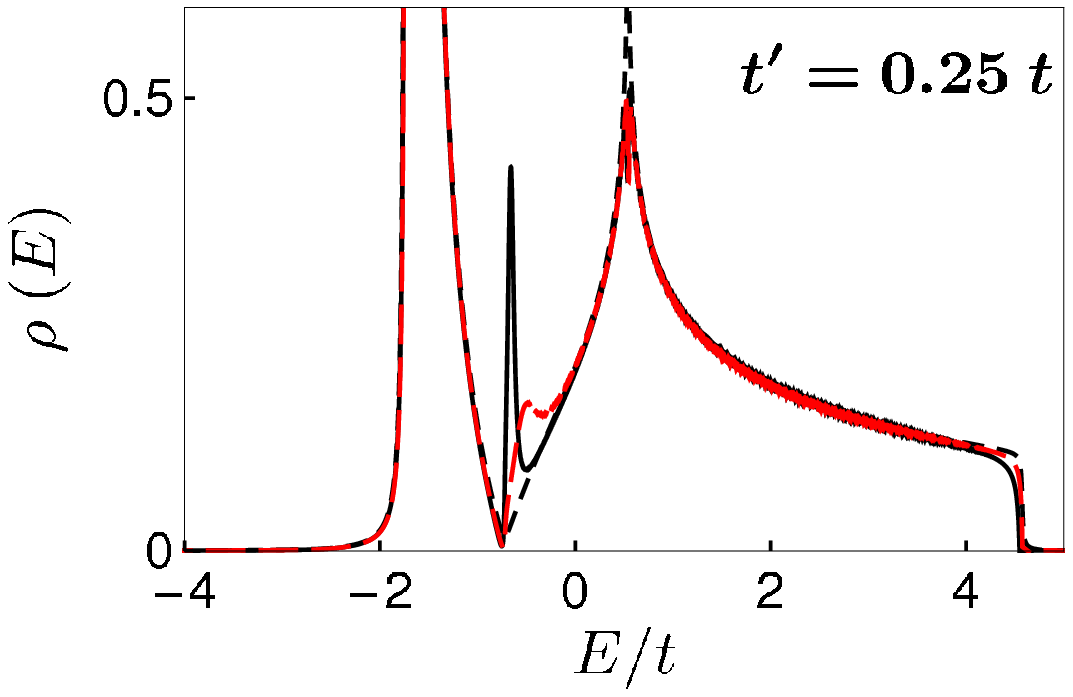}
\caption{ Evolution of the total DOS $\rho (E)$ with the strength of the impurity potential: $U_0/ t = 0$ (black dashed line), $5$ (black solid line), and $100$ (red dashed line), as obtained from Eq. (\ref{DOSTot}) ($N=20$ was used in the figures).
}
\label{tot-DOS}
\end{figure}

  As shown by Pereira et al.  \cite{castroneto}, the zero-mode state resides on the B sublattice alone, if only the $NN$ interaction is present. This is clearly seen from the top part of Fig. \ref{sublattice-DOS} and will be further discussed later when we discuss the Lippmann-Schwinger wave function. The emergence of the
 zero-mode state as the impurity potential is gradually increased to infinity is shown in Fig. (\ref{tot-DOS}).

\begin{figure}
\centering
\includegraphics[scale=0.6]{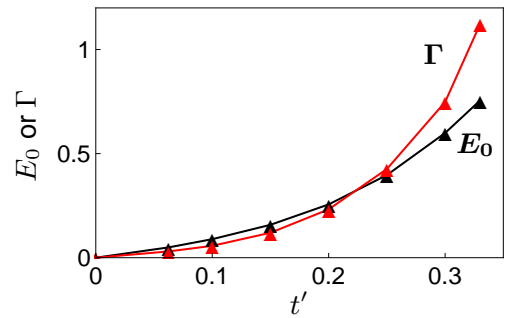}
\caption{ Energy of the resonance impurity state as measured with respect to the Fermi energy and the width of the resonance peak as obtained from the Lorentzian expression Eq. (\ref{Lorentz}). All quantities are in units of $t$.
}
\label{Lorentzian}
\end{figure}

Near resonance, the total DOS may be written as a Lonrentzian
\begin{equation}
\rho(E) = 2\rho_0(E)+\frac{1}{\pi N}\frac{\Gamma}{(E-E_0)^2+\Gamma^2},
\label{Lorentz}
\end{equation}
by expanding the real part of the GF: $F_0(E)=U_0^{-1}+F'_0(E_0)(E-E_0)+...$, valid near the
resonance energy. The width of the resonance peak is
$
\Gamma=- \pi \rho_0(E_0)/ F'_0(E_0),
$
which is non-zero if the $2NN$ interaction is present. The energy and width of the resonance peak is shown in Fig. \ref{Lorentzian} as a function of $t'$. The resonance peak disapperas for $t'  > \sim  0. 4 \ t$ or so.

\begin{figure}
\centering
\includegraphics[scale=0.65]{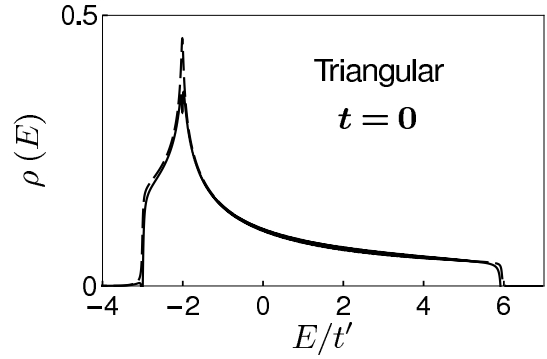}
\caption{ Unperturbed (red dashed line) and perturbed (black solid line) DOS for a vacancy in the triangular  lattice.
}
\label{Triangular-DOS}
\end{figure}

For the triangular lattice, as seen from the zeros of the real part of the on-site GF in Fig. (\ref{Fig-Dyson}), there is a resonance state at the
van Hove singularity at $E_0 = -2 t'$ and there is a bound state above the bands at the energy $E_0 \rightarrow \infty$ as
$U_0 \rightarrow \infty$ (the latter state is not shown in our figures). The unperturbed and perturbed DOS are shown in Fig. (\ref{Triangular-DOS}).
%

\section{Wave function of the impurity state}
As already mentioned, if only the $NN$ interaction is present in a bipartite lattice, a single vacancy in one sublattice (the minority sublattice now since there is one atom less) (a) introduces a state at the zero of energy (the onsite energy in the tight-binding model) and (b) this state furthermore lives on the majority sublattice alone. The first result follows from the particle-hole symmetry, which ensures the real part of the GF to be antisymmetric about the band center $E=0$, so that the infinite vacancy potential yields a state at that energy following  the  resonance condition obtained from the Dyson's equation. The second result naturally follows by examining the perturbed wave function in the presence of the impurity potential following the Lippmann-Schwinger equation. From this equation, we also estimate the spread of the wave function to the minority sublattice if the $2NN$ interactions are present.

The Lippmann-Schwinger equation relates the perturbed wave function to the unperturbed wave function via the GF $| \Psi \rangle = | \Psi^0 \rangle  + G^0 V | \Psi \rangle$. Inverting it, we get
\begin{equation}
\Psi_{i \alpha} \equiv \langle i \alpha | \Psi\rangle = \Psi_{i \alpha}^0 +
\frac{U_0 G^0_{i\alpha , 0A} (E) \Psi_{0 A}^0}{1- U_0 G^0_{0A , 0A} (E)}
\label{Psi-LS-0}
\end{equation}
and, using the resonance condition Eq. (\ref{E0}),  obtain the result for the
resonance state
\begin{equation}
\Psi_{i \alpha} = \Psi_{i \alpha}^0 +i \Psi_{0 A}^0 \frac{G^0_{i\alpha , 0A} (E_0) }{\rm{Im} \ G^0_{0A , 0A} (E_0)}.
\label{LS1}
\end{equation}
To examine the perturbed wave function, we need to examine the low-energy behavior of the GF since $E_0 \rightarrow 0$.
\begin{figure}[t]%
\centering
\includegraphics[width=6.5cm]{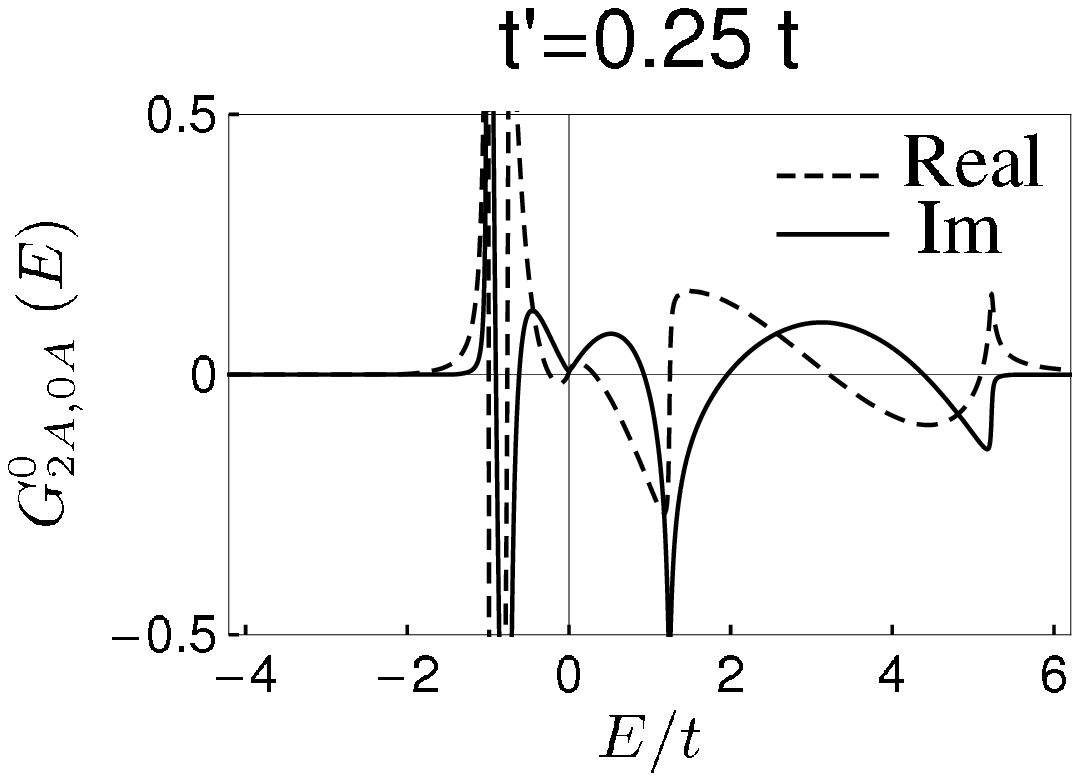}
\includegraphics[width=6.5cm]{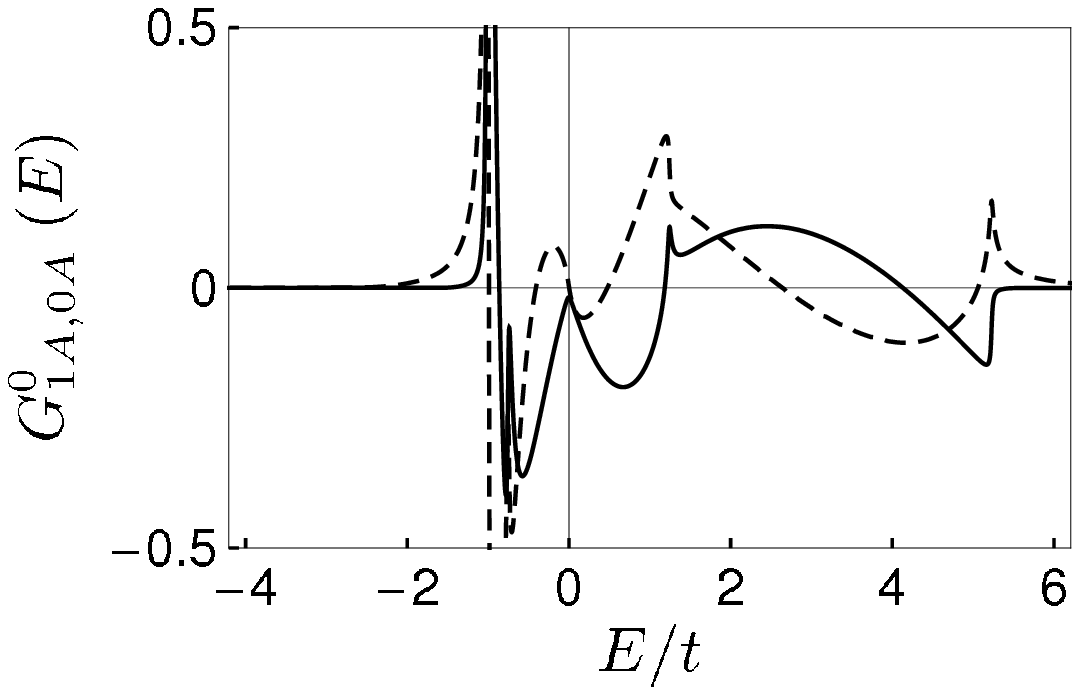}
\includegraphics[width=6.5cm]{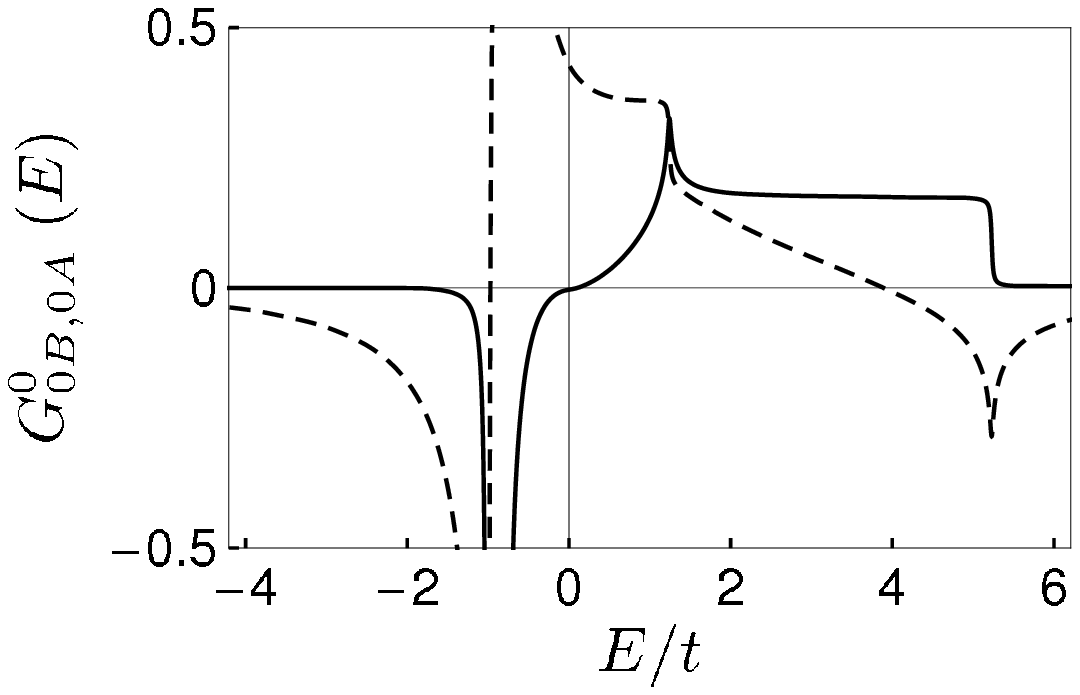}
\includegraphics[width=6.5cm]{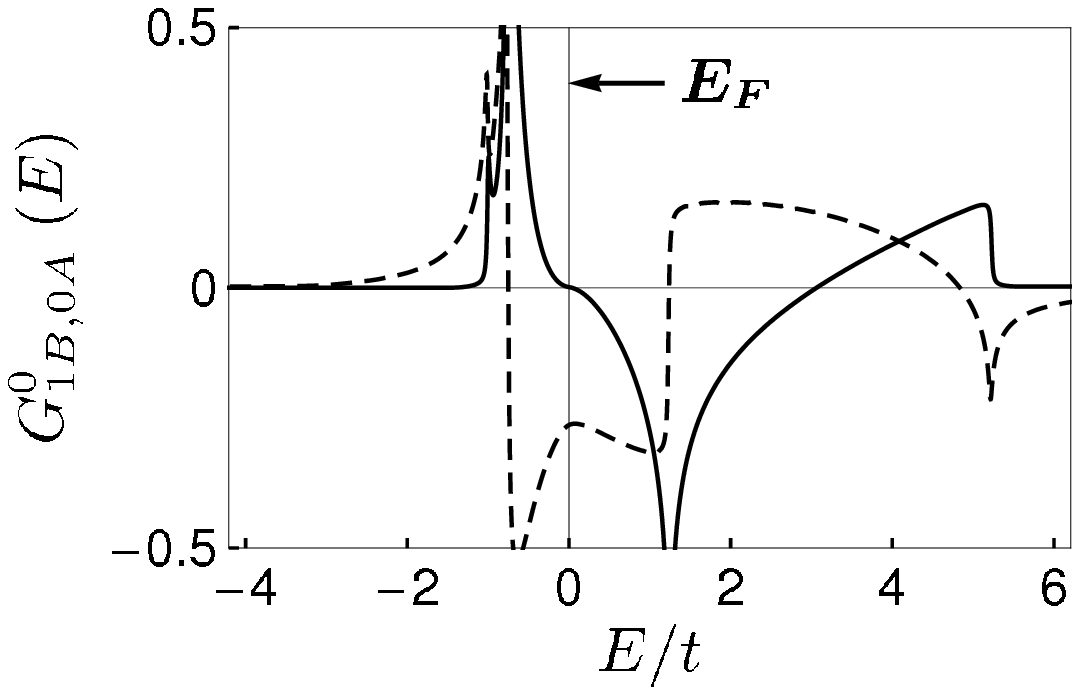}
\caption{ Calculated GF with the $2NN$ interaction present. Real and imaginary parts are shown by dashed and solid lines, respectively. Energy scale has been shifted by $3 t'$ so that the zero coincides with the Fermi energy, which continues to occur at the Dirac points for $t' < t/3$. Results for $G^0_{0A, 0A}$ were shown in the middle part of Fig. \ref{Fig-Dyson}, except that in that figure the zero of the energy was not shifted.
}
\label{G-tprime}
\end{figure}

{\it Resonance-state wave function for $t' = 0$}:   This case was discussed in detail in our earlier work   \cite{Ranjit-defect}; here we briefly summarize the results. For the tight-binding Hamiltonian with only $NN$ interactions, the low-energy GFs are given by
\begin{eqnarray}
G^0_{i A, 0A} (E) &=&\frac{A_c  \alpha }{2 \pi v_F^2} \times
\begin{cases}
( E \ln \frac{|E|r}{2 v_F}
 - \frac{i\pi}{2} |E| ) & \text{ $i \ne 0$}, \nonumber   \\
( E \ln \frac{|E| r_0}{2 v_F}
 - \frac{i\pi}{2} |E| ) & \text{ $i = 0$},
\end{cases} \nonumber   \\
G^0_{i B, 0A} (E) &=&  \frac{A_c {\rm Im} \beta}  {2 \pi v_F^2}  \times (  -\frac{v_F}{r}  -  \frac{i r \pi}{4 v_F} E |E| ),
\label{GF-E}
\end{eqnarray}
where  only the lowest order terms in energy have been kept in both the real and the imaginary parts.
In Eq. (\ref{GF-E}), $r_0 \sim 0.6 a$, $A_c$ is the unit cell area,  and $\alpha$ and $\beta$ are numbers of the order of one. Plugging this into Eq. (\ref{LS1})  and keeping the dominant terms as $E_0 \rightarrow 0$, the perturbed wave function becomes
\begin{equation}
\Psi =
\left( \begin{array}{c}
\Psi_{iA} \\
\Psi_{iB}
\end{array}\right) \
=
\left( \begin{array}{c}
|E_0| \ln |E_0| \\
c_i
\end{array}\right)
\rightarrow
\left( \begin{array}{c}
0 \\
c_i
\end{array}\right),
\label{Psi}
\end{equation}
where $c_i$ is the real part of $G^0_{i B, 0A}$ in Eq. (\ref{GF-E}).
Eq. (\ref{Psi}) implies that the vacancy-induced ``zero-mode"  state resides only on the $B$ sublattice and its amplitude is proportional to the real part of the corresponding GF. Eq. (\ref{GF-E}) also contains the striking feature that the wave function decays slowly, only as $1/r$ with distance, as seen from the real part of the GF
$G^0_{i B, 0A}$.

{\it Resonance-state wave function for $t' \ne 0$}:
We discuss the case where $t' \le t/3$, so that the Fermi energy is still at the linearly dispersive Dirac point. For higher values of $t'$, linearity is destroyed and the resonant-state solution near $E_F$ quickly disappears.
When the $2NN$ interaction is present, the particle-hole symmetry is lost and the real part of the on-site GF
is  no longer zero at the Fermi energy. The imaginary part of the on-site GF is still zero since the band structure remains linear with a vanishing DOS at the  Fermi energy which occurs at $-3 t'$.

The low energy expressions for the behavior of the GF are needed to examine the wave function of the resonance state. Unfortunately, it is not easy to derive an analytical expression for the GF similar to Eq. (\ref{GF-E}) for the
full tight-binging bands with the $2NN$ interactions present. Instead, we have studied numerically the behavior of the GFs  (some typical GFs are shown in Fig. \ref{G-tprime}) and extracted the low-energy behavior, which goes as
\begin{eqnarray}
G^0_{0A, 0A} (E) &=&a_0 + b_0 E + ic_0 |E|,   \nonumber   \\
 G^0_{iA, 0A} (E) &=&a_i E + ib_i |E|,   \hspace{2.0cm} \text{ $i \ne 0$}, \nonumber   \\
G^0_{i B, 0A} (E) &=& d_i + i e_i E,
\label{GF-E2}
\end{eqnarray}
where only  the lowest order terms have been kept in both the real and the imaginary parts and energy is measured from the Dirac point energy $-3t'$, which is the same as $E_F$ also.

From the low-energy behavior of the GF, Eq. (\ref{GF-E2}), we can infer that the resonance state wave function spreads into both sublattices now.
To see this, we insert  Eq. (\ref{GF-E2}) into the Lippmann-Schwinger Eq. (\ref{LS1}) and using the unperturbed wave function Eq. (\ref{estate}),  we get the result for the perturbed wave function:
$\Psi_{iA} = (2N)^{-1/2} e^{i \theta_{\bm{k}}   }   [ e^{i\bm{k} \cdot \bm{r}_{iA}  } + i c_0^{-1} (a_i+i b_i)]$ and
$\Psi_{iB} = (2N)^{-1/2} [ e^{i\bm{k} \cdot \bm{r}_{iB} } + i e^{i \theta_{\bm{k}}  }   c_0^{-1} (d_i/E_0+i e_i)]$. For small $t'$, the resonant energy $E_0 \propto t'$ (see Fig. \ref{Lorentzian}), so that the dominant part of the wave function is the second term of $\Psi_{iB}$. This leads to the result
$\Psi_{iA}^2/ \Psi_{iB}^2  \sim c_0^{2} d_i^{-2}  E_0^2 \sim (t'/t)^{2}$,
where we have used the rough behavior $c_0 \sim 1/t^2$ and $d_i \sim 1/t$. Thus the wave function spreads to the minority $A$ sublattice in the presence of the $2NN$ interaction roughly as
\begin{equation}
\Psi =
\left( \begin{array}{c}
\Psi_{iA} \\
\Psi_{iB}
\end{array}\right) \
=
\left( \begin{array}{c}
\alpha ( t'/t) \\
1
\end{array}\right).
\label{Psi2}
\end{equation}
This is consistent with Eq. (\ref{Psi}) that the wave function resides on the majority $B$ sublattice alone, when the $2NN$ interaction is absent.

\begin{figure}
\centering
\includegraphics[scale=0.6]{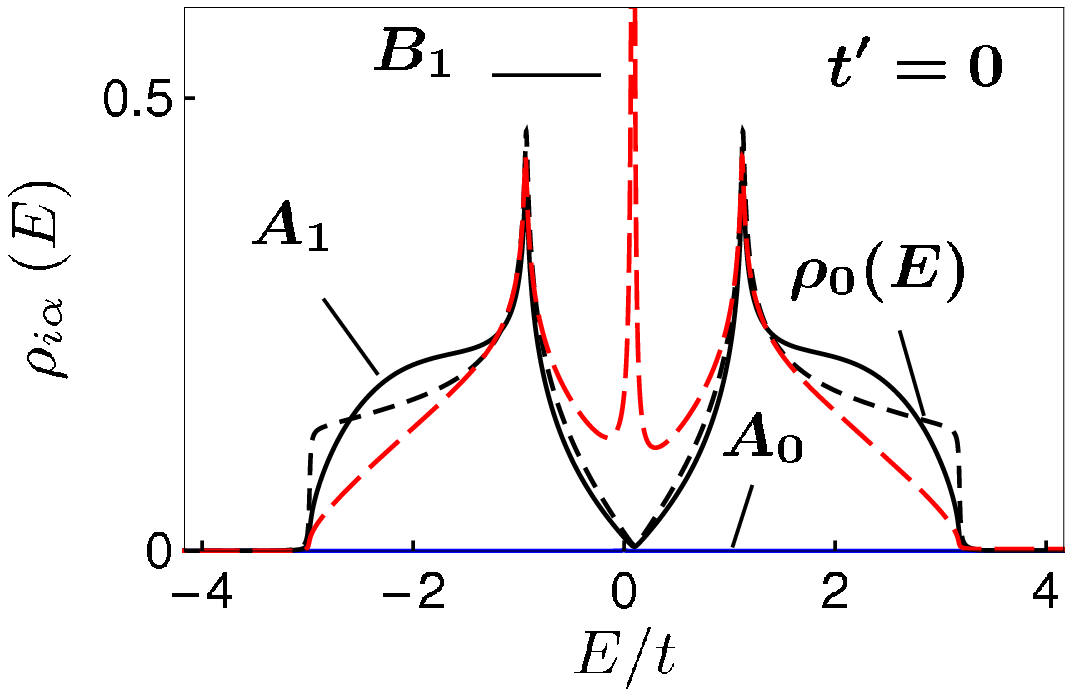}
\includegraphics[scale=0.6]{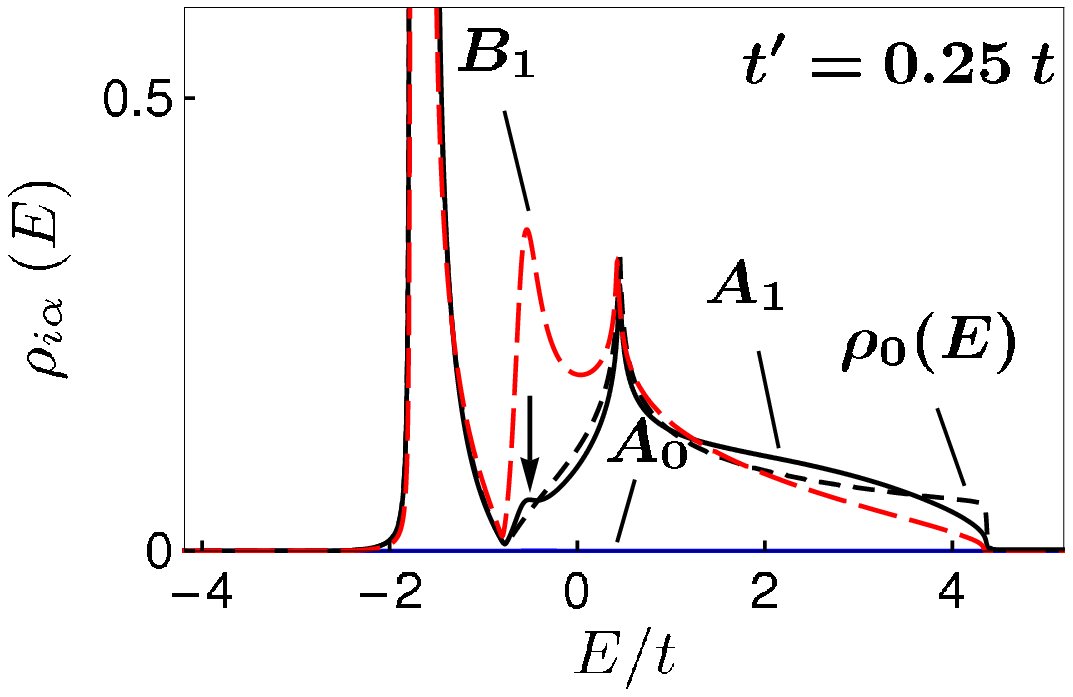}
\caption{ Local DOS at individual sites labelled in Fig. (\ref{graphene}) without (top) and with (bottom) the $2NN$ hopping. Unperturbed DOS $\rho_0 (E)$ is shown by the black dashed lines and $U_0$ was taken to be $100 t$. Top figure indicates that there is no contribution to the zero-mode state from the $A$ sublattice, while in the bottom figure, the $A_1$ site does have a small contribution (shown by the arrow).  Clearly, the zero-mode state which lives on the $B$ sublattice, if only the $NN$ hopping is present, has a strong component from the $B_1$ site in both cases shown. The contribution of the $A_0$ site is of course zero in the limit $U_0 = \infty$.
}
\label{LDOS-zeromode}
\end{figure}

In Fig. (\ref{LDOS-zeromode}), we show the site-specific local DOS for sites near the vacancy with the sites labelled in Fig. (\ref{graphene}). The figure indicates that the wave function for the zero-mode state has contributions only from the majority $B$ sublattice if $t' = 0$ and there is contribution from both the sublattices if $t' \ne 0$.

\section{Summary and Discussions}

In summary, we studied the formation of the impurity states in the honeycomb lattice, where both the $NN$ and the $2NN$ interactions are present. In the limit of large $2NN$ interaction, the honeycomb lattice  effectively goes over to two decoupled triangular lattice. If the $2NN$ interaction is absent, the impurity in the vacancy limit (impurity potential $\rightarrow \infty$) leads to the well-known zero-mode resonance state, introducing a $\delta$-function peak in the DOS. As the $2NN$ interaction is increased, the resonance state turns into a broadened peak, but quickly disappearing altogether for $t' > \sim 0.4 t$. For $t' = 0$, the resonance state is completely confined to the majority $B$ sublattice, i. e., opposite to the sublattice in which the vacancy is present, while with $t' \ne 0$, the resonance-state wave function spreads into both sublattices.

For the case of graphene, the $2NN$ interaction is small with $ t' \sim 0.06 t$, so that the zero-mode state occurs as a sharp peak with energy just above the band center energy $E=0$. An estimate using the actual hopping parameters given earlier and results of Fig. (\ref{Lorentzian}) indicates the peak position to be $E_0 \approx 0. 11$ eV and a similar peak width $\Gamma \approx 0.08$ eV. In graphene, there are also the $sp^2\sigma$ dangling bond electrons on the three carbon atoms adjacent to the vacancy. Three electrons occupy these states and the sole remaining electron occupies the zero-mode state discussed here. The strength of the Hund's coupling between the $\sigma$ and the $\pi$ electrons is about $0.4$ eV, which would pull down the energy of the zero-mode state by $0.2$ eV or so. This means that the zero-mode state becomes spin-polarized with one spin state occupied and the other empty, leading to a magnetic center. Density-functional calculations show that this is indeed the case and the vacancy produces a magnetic center with the vacancy-induced zero-mode $\pi$ band state half filled.
Recently, experimental evidence using scanning tunneling microscopy has been obtained for the existence of the zero-mode state  \cite{PRL2010}.
We also note that irrespective of the position of the resonance state, which occurs never too far away from $E_F$,
it is unlikely for it to be filled by two electrons considering the Coulomb interaction, so that the resonance state is filled by one electron and remains magnetic.

\begin{acknowledgement}
This work was supported by the U. S. Department of Energy through Grant No.
DOE-FG02-00ER45818.
\end{acknowledgement}



\begin{thebibliography}{[1]}

\bibitem{Review1} A. H. Castro Neto, F. Guinea, N. M. R. Peres, K. S. Novoselov, and A. K. Geim, Rev. Mod. Phys. {\textbf 81}, 109 (2009)

\bibitem{Review2} D. S. L. Abergel, V. Apalkov, J. Berashevich, K. Ziegler, and T. Chakraborty, Adv. Phys. {\textbf 59}, 261 (2010)

\bibitem{castroneto} V. M. Pereira, J. M. B. Lopes dos Santos, and A. H. Castro Neto, Phys. Rev. B {\textbf 77}, 115109 (2008)

\bibitem{hjort} M. Hjort and S. Stafstr\"om, Phys. Rev. B {\textbf 61}, 14089 (2000)

\bibitem{Ranjit-defect} B. R. K. Nanda, M. Sherafati, Z. Popovi\'c, and S. Satpathy, Phys. Rev. B (to be submitted)

\bibitem{Sankey} O. F. Sankey and J. D. Dow, Phys. Rev. B {\textbf 27}, 7641 (1983); O.F. Sankey, J. D. Dow, and K.   Hess, Appl. Phys. Lett. {\textbf 41}, 664 (1982)

\bibitem{Nanda-Graphene}%
 B. R. K. Nanda and S. Satpathy,
 Phys. Rev. B \textbf{80}, 165430 (2009)

 \bibitem{Horiguchi}%
T. Horiguchi,
 J. Math. Phys. \textbf{13}, 1411 (1972)

 \bibitem{M-RKKY}
 M. Sherafati and S. Satpathy, Phys. Rev. B (2011, in press)



\bibitem{PRL2010} M. M. Ugeda, I. Brihuega, F. Guinea, and J. M. Gomez-Rodriguez,
Phys. Rev. Lett. \textbf{104}, 096804 (2010)






\end{thebibliography}
\end{document}